**Surface Reduction State Determines Stabilization and Incorporation of Rh on α-Fe$_2$O$_3$(1$\bar{1}$02)**

*Florian Kraushofer, Nikolaus Resch, Moritz Eder, Ali Rafsanjani-Abbasi, Sarah Tobisch, Zdenek Jakub, Giada Franceschi, Michele Riva, Matthias Meier, Michael Schmid, Ulrike Diebold, Gareth S. Parkinson\**

F. Kraushofer, N. Resch, A. Rafsanjani-Abbasi, S. Tobisch, Dr. Z. Jakub, Dr. G. Franceschi, Dr. M. Riva, Dr. M. Meier, Prof. M. Schmid, Prof. U. Diebold, Prof. G. S. Parkinson
Institute of Applied Physics, TU Wien, Wiedner Hauptstraβe 8-10/E134, 1040 Wien, Austria
E-mail: parkinson@iap.tuwien.ac.at

M. Eder
Chair of Physical Chemistry, TU München, Lichtenbergstraße 4, 85748 Garching, Germany

Dr. M. Meier
Faculty of Physics and Center for Computational Materials Science, University of Vienna, Kolingasse 14-16, 1090 Wien, Austria

Iron oxides (FeO$_x$) are among the most common support materials utilized in single atom catalysis. The support is nominally Fe$_2$O$_3$, but strongly reductive treatments are usually applied to activate the as-synthesized catalyst prior to use. Here, we study Rh adsorption and incorporation on the (1$\bar{1}$02) surface of hematite (α-Fe$_2$O$_3$), which switches from a stoichiometric (1 × 1) termination to a reduced (2 × 1) reconstruction in reducing conditions. Rh atoms form clusters at room temperature on both surface terminations, but Rh atoms incorporate into the support lattice as isolated atoms upon annealing above 400 °C. Under mildly oxidizing conditions, the incorporation process is so strongly favoured that even large Rh clusters containing hundreds of atoms dissolve into the surface. Based on a combination of low energy ion scattering and scanning tunnelling microscopy data, as well as density functional theory, we conclude that the Rh atoms are stabilized in the immediate subsurface, rather than the surface layer.

## 1. Introduction

Stabilizing ever smaller metallic clusters on inexpensive metal oxide supports has been a long-standing goal of catalysis research. So-called "single-atom" catalysts (SACs) represent the ultimate limit of this endeavour, but stabilizing single atoms against agglomeration under reaction conditions is challenging.[1-7] To be stable, the metal atoms must form chemical bonds with the support, which affects their electronic structure and catalytic properties. To accurately model such a system requires the atomic-scale structure of the active site to be known. This information is extremely difficult to ascertain using current experimental methods, particularly under reaction conditions, and most theoretical calculations assume a high-symmetry site on an idealized periodic surface.[1, 8-11] Transmission electron microscopy (TEM) images usually suggest that isolated adatoms are located in cation-like sites relative to the bulk structure,[1, 12-17] but it is important to realise that, with this technique, neither the termination of the support nor the binding coordination of the adatom can be unambiguously determined. Some information is often inferred from ex-situ x-ray absorption spectroscopy (XAS) measurements, but these area-averaging methods do not necessarily probe the active sites on inhomogeneous samples, and rely on comparison of the SAC sample to standard samples such as a metal oxide, which may have very different coordination environments. In any case, the reactive environment almost certainly changes the state of the support surface, and with it the coordination environment of the single atom sites. In-situ TEM and XAS are possible for such systems,[18] but are not commonly applied.[19]

Iron oxides ($FeO_x$) are popular support materials in heterogeneous catalysis because they are inexpensive and chemically robust. In SAC studies, the as-synthesized support is nominally hematite ($\alpha$-$Fe_2O_3$). Consequently, the $\alpha$-$Fe_2O_3$(0001) surface with or without point oxygen vacancies is often chosen to represent the $FeO_x$ support in SAC studies.[1, 8-11] However, the atomic-scale structure of $\alpha$-$Fe_2O_3$(0001) remains particularly controversial, even under highly controlled ultrahigh vacuum (UHV) conditions.[20] There is evidence for a ferryl termination in oxidising conditions,[21, 22] and a whole host of long-range Moiré reconstructions occur when the surface becomes reduced.[20, 23] Eventually, when reduced more, the surface transforms to $Fe_3O_4(111)$.[24] Recently, we have studied the α-$Fe_2O_3(1\bar{1}02)$ surface, which is non-polar and similarly prevalent as the (0001) orientation on nanomaterial.[20, 25] This surface exhibits two terminations: a simple, stoichiometric (1 × 1) termination, and a reduced (2 × 1) reconstruction containing surface $Fe^{2+}$.[26, 27]

In this paper, we explore how Rh adsorbs on these surface terminations, and if stabilization is affected by oxidation and reduction of the iron oxide support surface. Using scanning tunnelling microscopy (STM), x-ray photoelectron spectroscopy (XPS) and low-energy $He^+$ ion scattering (LEIS), we find that Rh forms small clusters on both surface terminations upon vapor-deposition at room temperature (RT), indicative of a low surface diffusion barrier. Upon annealing, however, Rh atoms are accommodated in the subsurface on both surfaces. On the stoichiometric (1 × 1) termination, Rh is stabilized as single atoms in the second cation layer when the sample is heated to 400 °C, and does not diffuse further into the bulk. On the reduced (2 × 1) termination, higher coverages of Rh initially sinter to nanometer-sized clusters upon annealing in UHV, and these then dissolve into the hematite lattice with mild oxidation at 520 °C. These observations imply a strong driving force for Rh incorporation into the immediate subsurface of hematite, indicating a potential route for redispersion of sintered particles.

## 2. Experimental Results

First, we investigated the stability of small amounts of Rh on α-$Fe_2O_3(1\bar{1}02)$-(1 × 1) by depositing 0.025 ML Rh on the freshly prepared surface at room temperature. In STM [**Figure 1**(a)], we clearly observe small Rh clusters on an otherwise pristine surface. Based on the expected number of Rh atoms per unit area after deposition, we can estimate that, on average, a cluster consists of 3-4 atoms. **Figure 2**(a) shows XPS results for the Rh 3$d$ region after deposition, and after consecutive heating steps at elevated temperatures. The samples were cooled to room temperature to acquire XPS and STM data between the heating steps. Additional STM images from the annealing series are shown in **Figure S1**. Immediately after deposition, the Rh 3$d$ peak is relatively broad, with a maximum at ≈307.8 eV [black line in Figure 2(a)]. Since the Rh 3$d_{3/2}$ peak always exactly mirrors the 3$d_{5/2}$ peak with a shift of 4.8 eV and two thirds of the intensity, we will only discuss the 3$d_{5/2}$ peak from here on. When fitting the Rh 3$d$ region, appropriate additional components, fully constrained in position, area and FWHM, were added for the 3$d_{3/2}$ peak.

After heating to 200 °C, the Rh 3$d_{5/2}$ peak sharpens and shifts to lower binding energies [yellow line in Figure 2(a)]. This correlates with the formation of larger clusters in STM [Figure S1(b)]. At higher temperatures, the peak splits into a component at the position associated with the larger clusters at ≈307.5 eV, close to that of metallic bulk Rh,[28] and a component at higher binding energy. The evolution of the area ratio between these two components is shown in Figure 2(c) (black). For simplicity, we refer to the second component as the "oxidized" contribution, though it should be noted that for the small clusters found initially, the shift could also be explained at least in part by final state effects related to the cluster size.[29, 30] Taking a possible asymmetry of the metallic peak into account, with a tail towards higher binding energy, would further reduce the fraction of oxidized Rh, so the percentages of metallic Rh shown in Figure 2(c) should be taken as a lower limit. After annealing to 550 °C, the metallic component disappears and the Rh 3$d_{5/2}$ peak sharpens significantly [red line in Figure 2(a)]. In this state, the peak contains a single component at 309.3 eV, which can be assigned to an oxidized Rh species. This binding energy is significantly higher than the ≈308.6 eV that were previously reported for bulk $Rh_2O_3$,[28, 31, 32] and is closer to that reported for $RhO_2$.[31] In STM, this development corresponds to a disappearance of the clusters, and single bright features are instead observed at negative sample bias [Figure 1(b)]. Some of these bright features coexisting with clusters are observed in STM starting at 300 °C [Figure S1 (c-f)], in good agreement with the appearance of the oxidized peak at 309.3 eV in XPS [Figure 2 (a)]. We attribute this to Rh being liberated from the clusters and substituted into Fe sites in the hematite surface. Some of these features are still grouped together locally after annealing to 500 °C, suggesting that the incorporated Rh remains in the vicinity of the original cluster due to slow diffusion within the hematite lattice. However, after annealing for longer times, a random distribution can also be achieved, as shown in **Figure S2**. Interestingly, we found no further decrease in the XPS peak intensity even after annealing for two hours at 520 °C in $2 \times 10^{-6}$ mbar $O_2$, which indicates that Rh does not diffuse further into the bulk under these conditions.

In the annealing series shown in Figure 1 and Figure S2(a-h), the consecutive heating steps were all performed in UHV, which results in the surface being mostly reduced to the $(2 \times 1)$ termination after heating to 550 °C. Patches of the $(1 \times 1)$ structure remain only at step edges [Figure S1(h)]. This is the expected behaviour, as annealing in UHV at these temperatures is the normal preparation procedure for the $(2 \times 1)$ termination.[26] In the presence of Rh, however, small patches of the reduced $(2 \times 1)$ termination were found at step edges already after heating to 400 °C [Figure S1(e, f)], which is insufficient to form the $(2 \times 1)$ termination in the absence of Rh. The $(1 \times 1)$ termination could be restored everywhere by the final heating step at 520 °C in $2 \times 10^{-6}$ mbar $O_2$ [Figure S1(i)] without loss of Rh signal in XPS [Figure 2(a), pink]. Qualitatively, this is the same end result as was obtained when heating in $2 \times 10^{-6}$ mbar $O_2$ directly after Rh deposition (Figure S2), which prevents the surface from being reduced in the first place.

On the α-$Fe_2O_3$($1\bar{1}02$)-$(2 \times 1)$ termination, the initial state immediately after Rh deposition is similar as on the $(1 \times 1)$ termination, with small clusters observed in STM [Figure 1(c)]. The behaviour of the Rh 3$d$ peak in XPS with increasing temperature also qualitatively resembles the trends observed on the $(1 \times 1)$ termination. The peaks were again fitted by a metallic and an oxidized contribution. However, analysis of the peak intensity ratios [Figure 2(c)] highlights some differences between the two cases: On the (1 × 1) surface, the metallic contribution initially increases up to 200 °C, corresponding to sintering without incorporation. On the (2 × 1) termination, the metallic component decreases from the very beginning, and disappears completely at 400 °C, at a lower temperature than on the (1 × 1) termination (500 °C). In STM, images taken after annealing to 300 °C [Figure 1(d)] resemble those of the pristine surface, with no visible signature of the Rh, although it remains present in XPS. Additional STM data from this annealing series is shown in **Figure S3**.

As the Rh $3d$ peak in XPS does not weaken significantly even at high annealing temperatures on either surface termination, and since Rh-related protrusions remain clearly visible in STM on the (1 × 1) surface, the obvious question is whether the Rh is located directly in the surface layer. To answer this, we performed LEIS with 1 keV He$^+$ ions on both terminations, an exquisitely surface-sensitive technique. The spectra (**Figure 3**) clearly show a Rh peak directly following Rh deposition, but no trace of Rh is found in LEIS after annealing. This strongly suggests that incorporated rhodium is situated in the subsurface, rather than the surface layer, for both the (1 × 1) and the (2 × 1) termination.

Finally, we explored how the findings for 0.025 ML of Rh extend to a higher coverage (0.1 ML). On the stoichiometric (1 × 1) surface, the results are essentially the same as for 0.025 ML, with all Rh atoms ultimately being incorporated into the surface. An STM image and the corresponding XPS data of 0.1 ML of Rh incorporated in the (1 × 1) surface is shown in Figure S2. On the reduced, (2 × 1)-reconstructed surface, however, we find a decidedly different behaviour for 0.1 ML Rh (**Figure 4**). In XPS [Figure 4(a)], the Rh $3d_{5/2}$ peak progressively shifts to 307.3 eV upon heating and remains there, in good agreement with literature values for metallic Rh.[28] Fits of the Rh $3d$ data are shown in **Figure S4**, using the same fitting procedure as described above for the low-coverage case. The metallic component dominates at all times, but an oxidized component is also present, most strongly after heating to 200 °C. In STM, this corresponds to sintering of Rh into large clusters, as shown in Figure 4(b) after annealing to 580 °C in UHV. The increased cluster height also explains the decrease of the overall Rh intensity to 49% of the value after deposition, since the photoelectron signal from Rh in deeper layers of the clusters is attenuated by the layers above. Additional STM images for the entire annealing series are shown in **Figure S5**. Assuming a spherical cap shape, we estimate that the cluster visible in Figure 4(b) contains about 1200 Rh atoms. This is likely an upper bound for the actual number of atoms, as the cluster shape is always convoluted with the tip shape in STM. Interestingly, even for these very large clusters, Rh atoms can still be completely dissolved and incorporated into the surface upon annealing in oxygen, as seen in the STM image in Figure 4(c). Again, bright features within the lattice are visible, and these are often agglomerated locally [orange arrows in Figure 4(c)]. Both STM and LEED [inset to Figure 4(c)] show that the surface itself is transformed from the reduced (2 × 1) reconstruction to the stoichiometric (1 × 1) termination, as expected after oxygen annealing. In XPS, the Rh $3d$ peak [pink line in Figure 4(a)] again shifts to 309.3 eV, as observed whenever Rh is incorporated into the hematite lattice [Figure 2(a,b)].

## 3. Density Functional Theory Calculations

To rationalize the apparent stabilization of Rh in the subsurface, we performed DFT calculations for the (1 × 1) termination, with one Fe atom substituted by Rh in a (2 × 2) supercell. Substitution was tested for the first eight cation layers, and relative energies are shown in **Figure 5**. We find the first subsurface layer (C2) to be energetically most favourable, with an energy gain of −0.49 eV compared to the immediate surface layer. The second subsurface layer (C3) is 0.14 eV worse than C2, and below that, the substitution energy is already converged to a bulk value (0.12 eV worse than C2). These energies explain why extended annealing of the system does not lead to diffusion of Rh into the bulk of the sample at the temperatures used in this work.

Substitution of multiple Fe atoms in layer C2 by Rh was tested in a (3 × 3) supercell. The substitution energy per Rh atom stayed the same when a second Rh atom was placed in either of the nearest-neighbour sites in the subsurface layer. This means that there is neither repulsive nor attractive interaction between neighbouring Rh atoms in these sites, and thus no enthalpic preference to accumulate or disperse them, consistent with the observation of areas with high Rh concentration in Figure 4(c). However, random dispersion is favoured to maximize entropy. In all tested configurations, the calculated spin magnetic moments for Fe suggest that all iron remained $Fe^{3+}$ (4 $\mu_B$), with no evidence of $Fe^{2+}$ (3.5 $\mu_B$).[26] No direct conclusions about the Rh charge state can be drawn from the spin magnetic moment of Rh, which we always found to be close to zero. However, the Rh ion in the preferred C2 configuration exhibits a Bader charge of only +1.15 e, lower than in $RhO_2$ (+1.45 e) and in $Rh_2O_3$ (+1.22 e), which suggests a formal charge state of 3+ or less. Based on the known overall charge of the slab and the fact that no iron appears to be reduced, we therefore assign the rhodium as $Rh^{3+}$.

## 4. Discussion

Our results show that, in contrast to $Fe_3O_4$(001),[33] neither the stoichiometric (1 × 1) surface nor the reduced (2 × 1) reconstruction of α-$Fe_2O_3$($1\bar{1}02$) stabilizes Rh adatoms at room temperature. However, on the (1 × 1) termination at least, single atoms can incorporate into the hematite lattice via a thermally activated process. This makes sense, because $Rh_2O_3$, the most stable Rh oxide, is isostructural with α-$Fe_2O_3$. Interestingly, our results show that even large clusters can be re-dispersed by very mild oxidation, and these atoms remain visible in STM [Figure 2 (b)]. However, the fact that no Rh signal remains in LEIS after incorporation (Figure 3) strongly suggests that the Rh atoms are in fact not situated in the topmost cation layer, but rather in the subsurface, in agreement with our DFT results. Since the presence of the incorporated Rh atoms is most clearly visible in filled-states images, where oxygen atoms are imaged,[26] we conclude that we observe a modified density of states for the surface oxygen atom that is bound directly to a Rh atom in the immediate subsurface layer, marked by a dashed circle in Figure 5 (b). The dopants are also visible at some positive bias values, as shown in **Figure S6**. This indicates that the electronic structure of surface Fe atoms in the vicinity of subsurface Rh is also modified, since we have shown previously that empty-states images show the surface iron atoms.[26]

The preference of Rh to assume a subsurface instead of a surface site can be understood in terms of its preferred oxidation state and its lower tolerance for undercoordinated environments compared to Fe. In its native oxides, Rh is always six-fold coordinated, either as Rh(III) in $Rh_2O_3$ or as Rh(IV) in $RhO_2$. Fe, on the other hand, appears as Fe(II) in both $Fe_3O_4$ and FeO, and has tetrahedral coordination to oxygen in the former. Crucially, the Rh octahedra in $Rh_2O_3$ have different preferred bond lengths than Fe,[34] which they can achieve more easily near the surface in the $Fe_2O_3$ lattice, as the lattice strain induced by substitution can be mitigated by surface relaxations. The Fe octahedra in bulk hematite are trigonally distorted to a $C_{3v}$ symmetry[25, 35] with two different bond lengths. However, in our DFT calculations, Rh in the first subsurface layer can realize its preferred, relatively uniform bond lengths of 2.05–2.08 Å,[34] as opposed to the hematite bulk Fe—O bond lengths of 1.97 Å and 2.12 Å, respectively. Together, these two effects favour the closest cation site to the surface in which a six-fold bonding environment can be achieved, which is in the first subsurface layer.

As to how the Rh is oxidized, there are two possible routes: When annealing in a background of oxygen, displaced iron likely diffuses to form more hematite at step edges or as new islands, as we have observed previously when depositing Ti (which also replaces Fe in layer C2).[36] When no gas-phase oxygen is available, small amounts of excess cations may be compensated without major reduction of the surface by diffusion of Fe into the bulk, as has been documented for $Fe_3O_4$.[20] However, it is worth noting that when depositing Rh on the (1 × 1) termination and annealing in UHV, we found small patches of the (2 × 1) termination sooner than expected, after heating to only 400 °C (Figure S1). This may be a compensation mechanism to help accommodate reduced Fe that has been displaced by Rh incorporation.

Concerning the oxidation state of incorporated Rh, the very high binding energy of 309.3 eV observed in XPS (Figure 2 and Figure 4) may indicate a $Rh^{4+}$ state, as the peak position for bulk $Rh_2O_3$ is usually given as ≈308.6 eV.[28, 32] However, in the absence of compensating O vacancies, this would require the reduction of iron to $Fe^{2+}$. We have shown previously that introducing $Ti^{4+}$ dopants into the surface induces a localized, oxidized restructuring of the surface, which allows all iron to remain $Fe^{3+}$.[36] Rh induces no such restructuring, and when annealing in oxygen to prevent partial reduction to the (2 × 1) termination, we do not observe any signature of $Fe^{2+}$ cations in grazing-emission XPS, even for 0.1 ML Rh [Figure S2(b)]. Furthermore, our DFT calculations show no evidence for reduction of $Fe^{3+}$ to $Fe^{2+}$ in the presence of Rh, and Bader charge analysis indicates a Rh charge state closer to that of Rh in $Rh_2O_3$. Based on these results, we assign the incorporated rhodium as $Rh^{3+}$, despite the unusually high binding energy in XPS.

Incorporation of Rh in a six-fold coordinated environment in the subsurface means that it would likely be catalytically inactive. Similar behaviour has been observed following calcination of Rh on anatase $TiO_2$ particles.[37] One can therefore see why reductive treatment is required to activate the catalysts following calcination.[37] However, the fact that lattice incorporation is favourable enough to abstract Rh even from very large clusters may hold interesting possibilities for regenerating single-atom catalysts. Typical deactivation mechanisms involve sintering and poisoning; both can be reversed if the catalyst is redispersed by an oxidation step (which also burns off carbonaceous species). Note, however, that we could not recover the Rh to the surface by reducing the (1 × 1)-terminated surface by UHV annealing, because Rh is also accommodated in the subsurface on the (2 × 1)-terminated surface. We have shown previously that Rh atoms are stabilized on $Fe_3O_4(001)$,[33] so perhaps an even stronger reduction of the surface could recover surface Rh species.

WILEY-VCHThe behaviour of Rh on the reduced (2 × 1) surface is different from that on the stoichiometric (1 × 1) termination. If we start from a 0.1 ML deposition, the reduced (2 × 1) termination retains large Rh clusters at temperatures where the stoichiometric surface has already taken all of the metal into the lattice (Figure S2 and Figure S5). This suggests that there is less energy gain for Rh incorporation on the (2 × 1) termination, and that the energy gained by incorporation of Rh is smaller than the cohesive energy in large, bulk-like Rh clusters. This also indicates that reduction should help with re-exposing the Rh atoms and activating them for catalysis. On the other hand, incorporation of small amounts of Rh appears to be more facile on the (2 × 1) surface, as the process begins at lower temperatures [Figure 2(c)]. Taking the apparently lower thermodynamic driving force into account, this suggests that the faster incorporation of low coverages on the (2 × 1) termination [Figure 2(c)] is due to lower diffusion barriers, perhaps due to the different, more open structure of the α-$Fe_2O_3$($1\bar{1}02$)-(2 × 1) reconstruction.

The qualitative difference between the low and high Rh coverage on the (2 × 1) termination can be explained by two different effects. On the one hand, since incorporating Rh requires it to be oxidized, Fe needs to be reduced simultaneously. Rh is more electronegative than Fe, so it is clear that one can only reduce Fe from 3+ to 2+ (which are both common oxidation states of Fe), not any further. Thus, for each incorporated Rh, three Fe atoms have to be reduced to $Fe^{2+}$. Unlike the stoichiometric (1 × 1) surface, the (2 × 1) reconstruction already contains large amounts of $Fe^{2+}$.[26] Rh incorporation will therefore be unfavourable unless diffusion of oxygen from the bulk to the surface, or of excess Fe to the bulk, is fast. If bulk diffusion is slow, displacing Fe upon Rh incorporation becomes increasingly harder. This would offset the energy gained by Rh incorporation, and would lead to a saturation behaviour. On the other hand, clusters may never reach a critical size in the low-coverage case (Figure S3), and thus never become thermodynamically stable against incorporation. Most likely, both effects need to be taken into account to correctly describe the behaviour of Rh during the annealing series performed here. However, insufficient bulk diffusion should become less relevant at higher temperatures. We can therefore conclude that once very large clusters are formed, as was the case for our experiments with 0.1 ML Rh, an equilibrium state is reached in which the clusters are indeed thermodynamically stable against incorporation.

These findings illustrate that it is important to account for the reconstructions formed under reducing conditions, instead of simply modelling the reduction by assuming that oxygen vacancies are introduced. In prior DFT studies, the α-$Fe_2O_3$(0001) surface was usually modelled using the $O_3$-termination capped with a layer of the catalyst atoms (equivalent to an Fe-termination with the terminating Fe layer completely substituted by the metal in question).[1, 8-11] In modelling CO oxidation, it has been assumed that an oxygen vacancy can be formed next to the SAC site, which can react with $O_2$ and CO, forming $CO_2$ and repairing the vacancy. To complete the cycle, another CO molecule reacts with a surface oxygen, forming a second $CO_2$ molecule and recovering the oxygen vacancy.[1, 8-11] However, the reduced α-$Fe_2O_3$(0001) surface exhibits an abundance of reconstructions,[20, 23] and removing O from an already reduced surface likely costs more energy. Of course, assuming a somewhat simplified model for the surface structure is necessary to make calculations tractable, but it seems unlikely that the real surface activated in reducing conditions would exhibit the bulk-truncated termination with easily available oxygen.

Furthermore, stabilization of single adatoms will depend strongly on both the oxidation state and the structure of the actual surface. Adatom stabilization depends more strongly on diffusion barriers than on pure binding energies, but generally, oxidized surfaces tend to bind cations more strongly, as we have seen on $Fe_3O_4$(001).[38] While defects can certainly act as trap sites for metal adatoms, oxygen vacancies have been reported to be ineffective at stabilizing metals such as Pt or Rh on $TiO_2$.[39-41] For covalently bound cations, this is not surprising, since the adatom loses a potential bond to the support, and bonding may be weakened further if charge transfer is inhibited by the presence of other reduced cations. A recent screening study of a wide range of different transition metals on α-$Fe_2O_3$(0001) reports strongly covalent binding for all of them.[10] Therefore, there is reason to assume that our finding of metal clusters forming on the reduced, but not on the oxidized surface may also be more generally applicable to $FeO_x$ surfaces.

Finally, for Rh specifically, it seems likely that the driving forces of preferred six-fold coordination on the one hand and lattice strain on the other hand can be generalized to other hematite surfaces, as well as to step defects. This may be helpful for stabilizing and re-dispersing catalyst atoms. A similar route has been demonstrated for Pt on ceria, where Pt can be abstracted from clusters into highly coordinated, but catalytically inactive sites under oxidizing conditions.[42] However, whether an atom is situated in the surface or in the first subsurface layer can be extremely hard to distinguish in imaging techniques such as STM [as seen in Figure 2(b)] or TEM, which is commonly used to identify atom positions in nanoparticle studies.[1, 12-17] Special care should therefore be taken to avoid erroneous identification of catalyst activity if the preparation leaves the single atom in an inactive subsurface site, or if it moves there under reaction conditions.

## 5. Conclusion

We have studied the interaction of Rh with α-$Fe_2O_3$(1$\bar{1}$02) both on the stoichiometric (1 × 1) and the reduced (2 × 1) termination. Neither surface stabilizes single Rh adatoms, and small Rh clusters are found after deposition at room temperature. Low coverages of Rh are incorporated in the substrate below 400 °C in both cases and are stabilized in the subsurface. Larger coverages of Rh sinter into clusters consisting of hundreds of atoms on the reduced (2 × 1) termination, but can be dissolved and re-dispersed by annealing in oxygen, which also transforms the surface back to the (1 × 1) termination. The incorporated oxidized Rh species substitute Fe in the first subsurface layer. We assign the features imaged with increased apparent height in STM to be the surface O atoms bound to the subsurface Rh.

**6. Methods**

All results presented in this work were collected in a UHV setup consisting of a preparation chamber (base pressure $< 10^{-10}$ mbar) and an analysis chamber (base pressure $< 5 \times 10^{-11}$ mbar). The system is equipped with a commercial low-energy electron diffraction (LEED) apparatus (VSI), a nonmonochromatic Al K$\alpha$ X-ray source (VG), an ion gun (He$^+$) used for LEIS, a SPECS Phoibos 100 analyser for XPS and LEIS, and an Omicron $\mu$-STM. The STM was operated in constant-current mode using electrochemically etched W tips. STM images were corrected for distortion and creep of the piezo scanner, as described in ref. [43]. Rh was deposited from an electron-beam evaporator (Omicron) in the preparation chamber. A quartz-crystal microbalance was used to calibrate the amount of deposited material, with deposition times of ca. 30–120 seconds for 0.025–0.1 monolayers (ML) of rhodium. Throughout this paper, we define a monolayer as the number of Fe atoms in the surface layer. 1 ML of Rh is therefore defined as two Rh atoms per $\alpha$-Fe$_2$O$_3$(1$\bar{1}$02)-(1 × 1) unit cell, which corresponds to a density of $7.3 \times 10^{14}$ atoms cm$^{-2}$.

The experiments were conducted on a single-crystalline, 0.03 at.% Ti-doped hematite film grown homoepitaxially by pulsed laser deposition on a natural $\alpha$-Fe$_2$O$_3$(1$\bar{1}$02) sample (SurfaceNet GmbH, 10 × 10 × 0.5 mm$^3$, <0.3° miscut), as described previously.[36] Doping was achieved by alternating deposition from an Fe$_3$O$_4$ single crystal target and a 1 at.% Ti-doped hematite target, home-synthesized from commercial TiO$_2$ and Fe$_2$O$_3$ powders (99.995% purity, Alfa Aesar). The mixed powders were pressed in an isostatic press at 400 MPa and room temperature in a cylindrical silicone mold, and sintered in an alumina-tube furnace (6 h at 1200 °C, 1 bar of flowing O$_2$, 5 °C/min ramp rates), as described in detail elsewhere.[44] The resulting hematite film is sufficiently conductive for STM at room temperature, with large atomically flat terraces. The surface appears identical to the undoped samples studied previously.[26] Before each experiment, the sample was re-prepared by sputtering (1 keV Ar$^+$ ions, ~2 μA) and annealing in oxygen (2 × 10$^{-6}$ mbar, 520 °C) for 30 min, which yields the stoichiometric (1 × 1) termination. For experiments on the reduced (2 × 1) reconstruction, the sample was then annealed in UHV at 580 °C for 15 min.

Density functional theory (DFT) calculations were performed using the Vienna *ab initio* simulation package[45, 46] (VASP) with the projector augmented wave method[47, 48]. The Perdew, Burke, and Ernzerhof exchange-correlation functional[49] was used together with a Hubbard $U$ ($U_{eff}$ = 4.0) to treat the highly correlated Fe 3$d$ electrons.[50] The same $U$ was applied for Rh atoms to avoid artificially biasing 3$d$ electron occupations among the different transition metal cations.[51] Rh substitution was tested in (2 × 2) supercells on asymmetric slabs consisting of 30 atomic layers, with the bottom five layers kept fixed, and using a 4 × 4 × 1 Γ-centred **k**-mesh. Coverage dependencies of substituting two Rh atoms in the same sublayer were tested in a larger (3 × 3) supercell, with an adjusted **k**-mesh of 3 × 3 × 1. The plane-wave basis-set cut-off energy was set to 450 eV, and convergence was achieved when residual forces acting on ions were smaller than 0.02 eV/Å. The charge states of Rh ions were evaluated using the Bader approach,[52-54] and benchmarked to calculations of bulk RhO$_2$ and Rh$_2$O$_3$ using the same computational setup.

**Supporting Information**
Supporting Information is available from the Wiley Online Library or from the author.

**Acknowledgements**
GSP, FK, ARA and MM acknowledge funding from the European Research Council (ERC) under the European Union's Horizon 2020 research and innovation program Grant agreement No. 864628. UD, FK and MR acknowledge support by the Austrian Science Fund (FWF, Z-250, Wittgenstein Prize). NR and ZJ were supported by the Austrian Science Fund (FWF, Y847-N20, START Prize). GF acknowledges support from the TU-D doctoral school of the TU Wien. The computational results presented were achieved using the Vienna Scientific Cluster (VSC 3).

Received: ((will be filled in by the editorial staff))
Revised: ((will be filled in by the editorial staff))
Published online: ((will be filled in by the editorial staff))

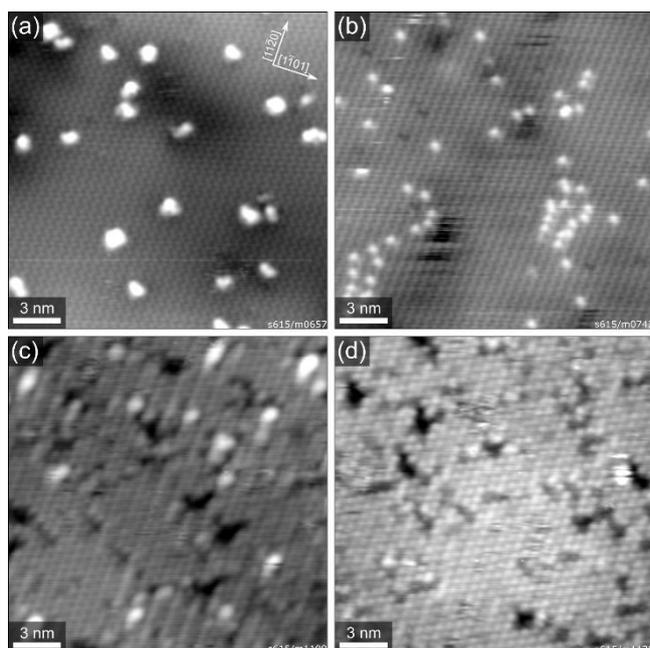

**Figure 1: STM images of 0.025 ML Rh on α-Fe$_2$O$_3$(1$\bar{1}$02).** (a) 0.025 ML Rh as deposited on the clean α-Fe$_2$O$_3$(1$\bar{1}$02)-(1 × 1) surface at room temperature ($U_{sample}$ = +3 V, $I_{tunnel}$ = 0.3 nA) and (b) after annealing at 500 °C for 15 minutes in UHV ($U_{sample}$ = −2.8 V, $I_{tunnel}$ = 0.1 nA). (c) 0.025 ML Rh as deposited on the clean α-Fe$_2$O$_3$(1$\bar{1}$02)-(2 × 1) surface ($U_{sample}$ = −3 V, $I_{tunnel}$ = 0.1 nA) and (d) after annealing at 300 °C for 10 minutes in UHV ($U_{sample}$ = −2.8 V, $I_{tunnel}$ = 0.1 nA). The images are from the same measurement series as the XPS results shown in Figure 2.

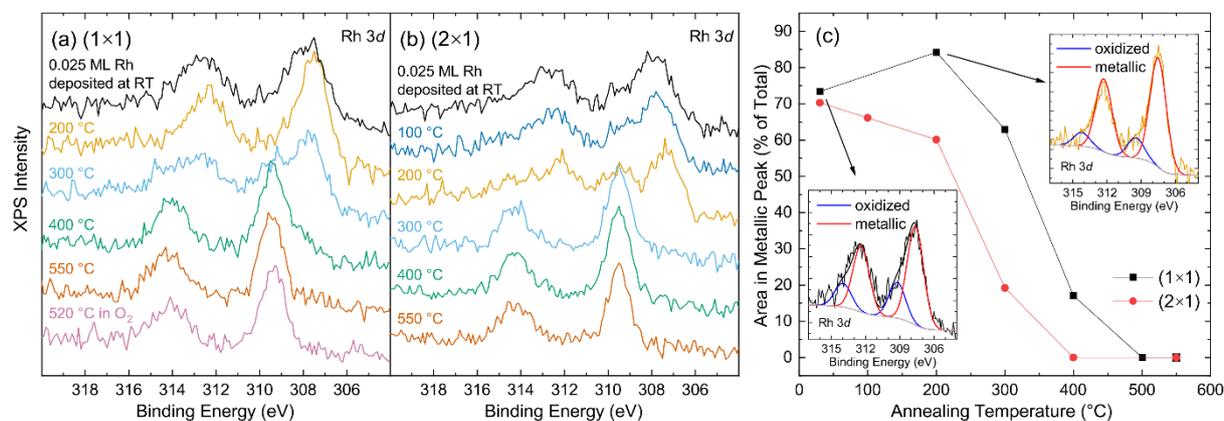

**Figure 2: Analysing thermal stability using the Rh 3d region in XPS.** (a,b): XPS spectra (Al Kα, 70° grazing emission, pass energy 16 eV, offset vertically for clarity) of the Rh 3d core-level peaks for 0.025 ML Rh on the α-Fe$_2$O$_3$(1$\bar{1}$02)-(1 × 1) and (2 × 1) surfaces, respectively, as deposited at room temperature and after successive annealing steps in UHV at different temperatures (at least 10 min per step). The spectra were acquired after cooling back to room temperature. For the bottom-most spectrum in (a), the sample was annealed for 30 min at 520 °C in a background of 2 × 10$^{-6}$ mbar O$_2$. (c) Area percentages for peak fits to the spectra in (a) and (b) using two components for Rh 3d$_{5/2}$. Fit results for the first two curves in (a) are shown in the insets to (c).

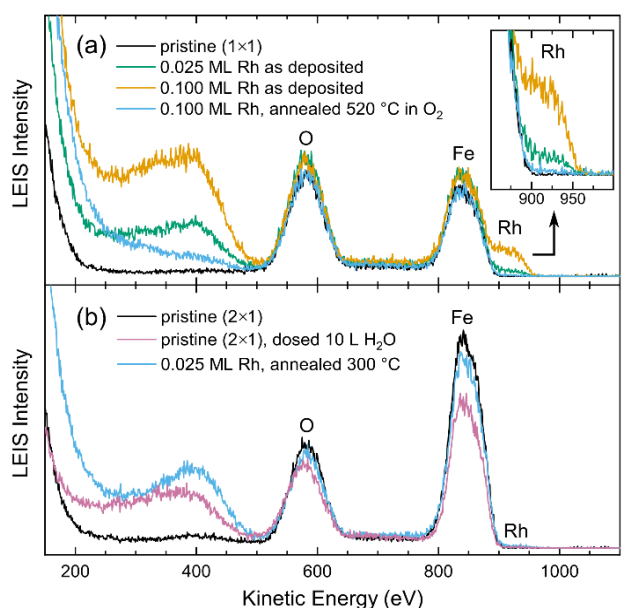

**Figure 3: He⁺ LEIS of as-deposited and incorporated Rh.** LEIS measurements (1 keV He⁺, 90° scattering angle) of (a) Rh on the α-$Fe_2O_3$(1$\bar{1}$02)-(1 × 1) surface and (b) on the (2 × 1) termination. In panel (a), as-deposited Rh is clearly visible as a peak at ≈920 eV for both the 0.025 ML and the 0.1 ML coverage. A magnified view of the Rh region is shown in the inset. The broad peak below 500 eV is correlated with adsorbed water, as demonstrated by the spectrum acquired after dosing 10 L $H_2O$ (1 L = $1.33 \times 10^{-6}$ mbar × s) on the clean (2 × 1) surface (b, pink). We attribute this peak to fast hydrogen recoils (at 90° scattering angle, these must be either H⁺ recoils deflected at surface atoms or H⁺ recoils created by He deflected at surface atoms).

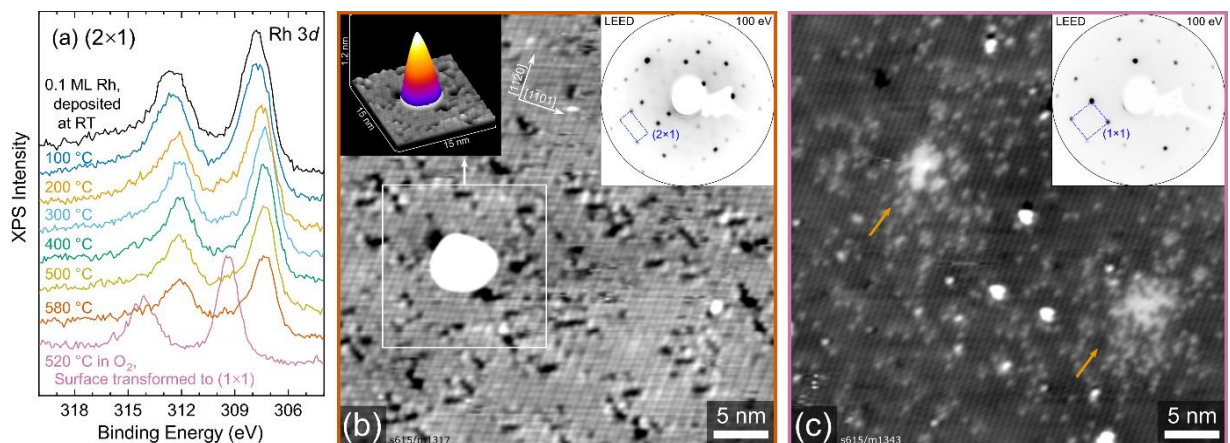

**Figure 4: 0.1 ML Rh on α-$Fe_2O_3$(1$\bar{1}$02)-(2 × 1).** (a) The Rh 3$d$ region in XPS (Al Kα, 70° grazing emission, pass energy 16 eV) after depositing 0.1 ML Rh on the α-$Fe_2O_3$(1$\bar{1}$02)-(2 × 1) surface, then successively annealed in UHV for 10 min at different temperatures. For the last spectrum (pink), the sample was annealed for one hour at 520 °C in a background of $2 \times 10^{-6}$ mbar $O_2$. All spectra were acquired after cooling the sample to room temperature. (b) 40 × 40 nm² STM image ($U_{sample}$ = −2.5 V, $I_{tunnel}$ = 0.1 nA) taken after heating the sample to 580 °C [red line in (a)]. Large clusters are found on the surface, one of which is plotted in 3D as an inset (ca. 1.1 nm apparent height). A LEED pattern of the surface with (2 × 1) periodicity is shown in the top right. (c) 40 × 40 nm² STM image ($U_{sample}$ = −3 V, $I_{tunnel}$ = 0.1 nA) taken after annealing the sample at 520 °C in $2 \times 10^{-6}$ mbar $O_2$ for 1 h. The surface periodicity has changed to (1 × 1), as also seen in the LEED pattern (inset). Orange arrows indicate local agglomerations of brighter features in the surface.

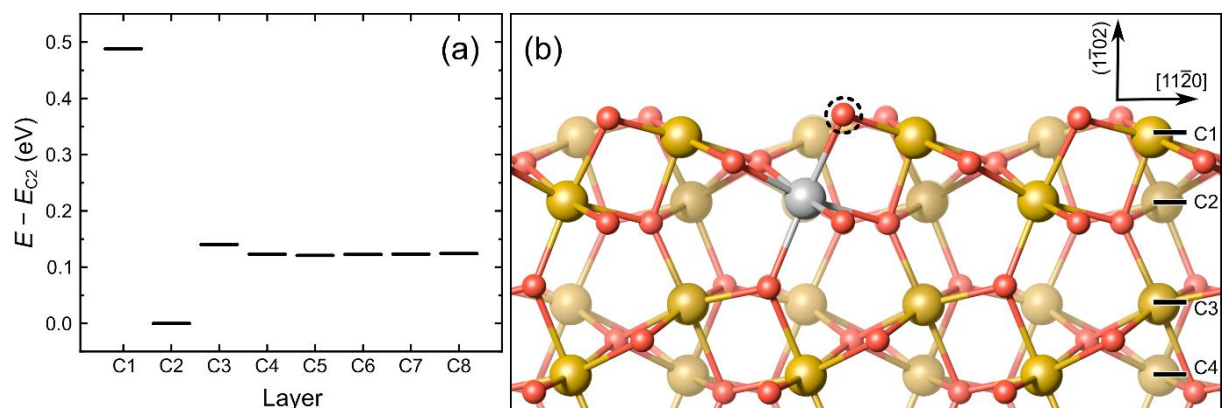

**Figure 5: DFT results for Rh substitution in the α-Fe$_2$O$_3$($1\bar{1}02$)-(1 × 1) surface.** (a) Substitution energies of one Rh atom replacing one Fe atom in a given layer, referenced to the substitution energy in the first subsurface layer (C2). (b) Side view (looking along the [$\bar{1}102$] direction) of the α-Fe$_2$O$_3$($1\bar{1}02$)-(1 × 1) surface, as in ref. [26]. Cation layers C1–C4 are labelled on the right. Iron is drawn as brown (large), oxygen as red (small) spheres. One iron atom in the first subsurface cation layer is replaced by Rh (grey), which corresponds to the most favourable substitution site according to panel (a). The surface oxygen atom bound directly to Rh is marked by a dashed circle. The direction perpendicular to the surface is labelled as ($1\bar{1}02$) in round brackets because there is no integer-index vector corresponding to that direction for the ($1\bar{1}02$) plane.

Adsorption of Rh is investigated on stoichiometric and reduced terminations of α-$Fe_2O_3$(1$\bar{1}$02). Neither surface stabilizes single Rh atoms at room temperature, but Rh incorporates as single atoms into the immediate subsurface at slightly oxidizing conditions. Indeed, this process is so favourable that even large clusters, consisting of hundreds of rhodium atoms, can be dissolved and re-dispersed in the surface.

F. Kraushofer, N. Resch, M. Eder, A. Rafsanjani-Abbasi, S. Tobisch, Z. Jakub, G. Franceschi, M. Riva, M. Meier, M. Schmid, U. Diebold, G. S. Parkinson*

**Surface Reduction State Determines Stabilization and Incorporation of Rh on α-$Fe_2O_3$(1$\bar{1}$02)**

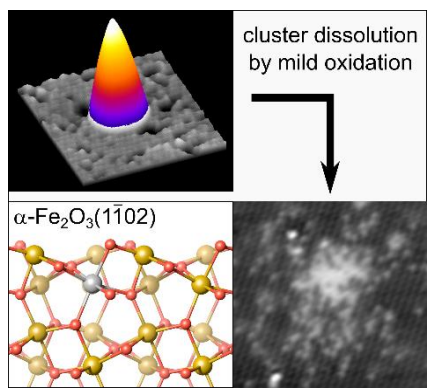

# Supporting Information

**Surface Reduction State Determines Stabilization and Incorporation of Rh on α-Fe$_2$O$_3$(1$\bar{1}$02)**

*Florian Kraushofer, Nikolaus Resch, Moritz Eder, Ali Rafsanjani-Abbasi, Sarah Tobisch, Zdenek Jakub, Giada Franceschi, Michele Riva, Matthias Meier, Michael Schmid, Ulrike Diebold, Gareth S. Parkinson\**

*\*Corresponding author: parkinson@iap.tuwien.ac.at*

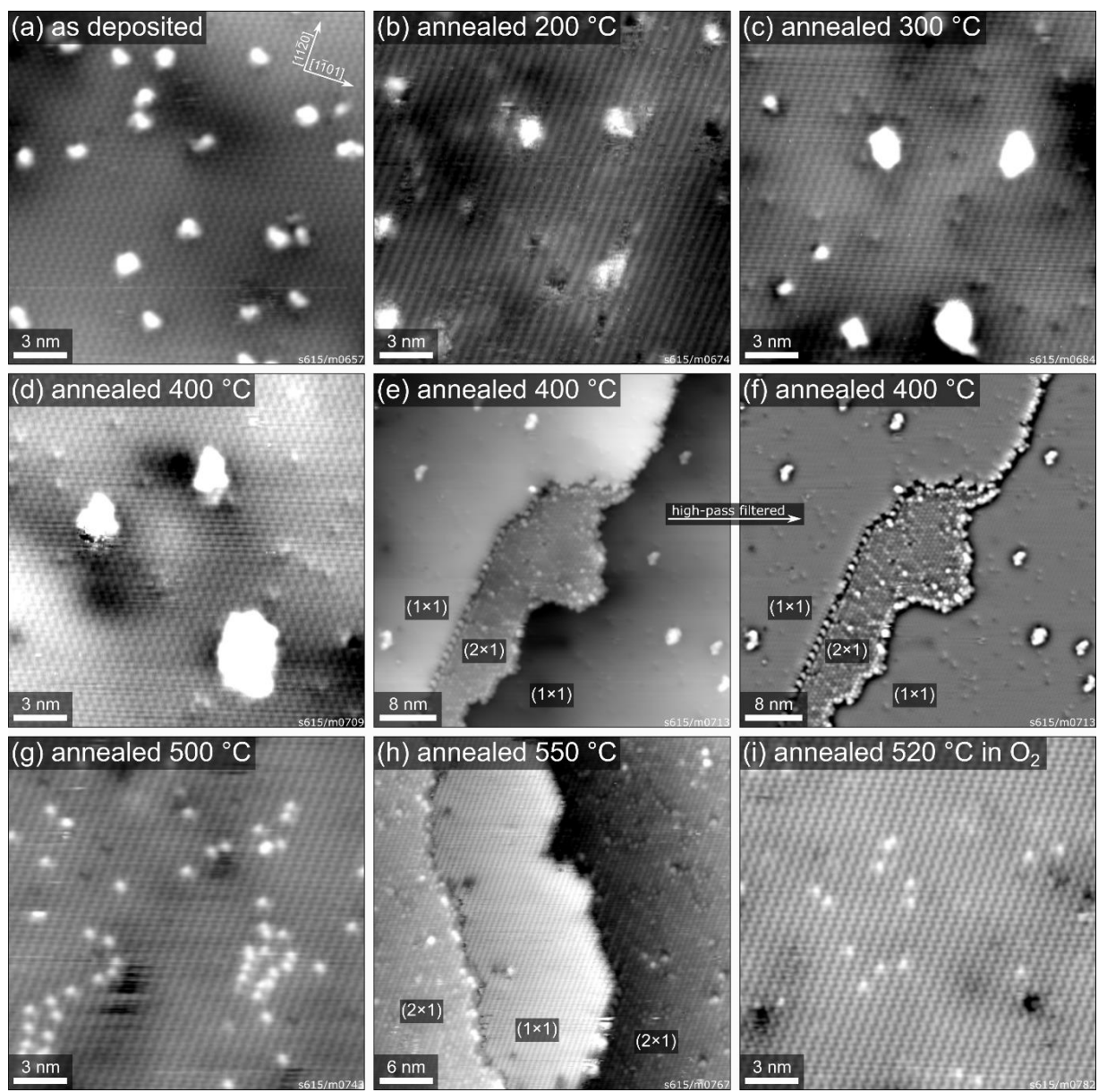

**Figure S1:** STM images corresponding to the XPS data in Figure 2 (a) of 0.025 ML Rh on α-Fe$_2$O$_3$(1$\bar{1}$02)-(1 × 1). Apart from panel (i), all annealing steps were performed in UHV. Panels (a) and (g) are the same as in Figure 1 (a, b). Panel (e) shows a region of interest in which a small patch of the reduced (2 × 1) termination was formed at a step edge after annealing to 400 °C. Panel (f) is the same image as (e), high-pass filtered for better visibility of the surface structure. In panel (h), the majority of the surface is (2 × 1)-terminated after annealing to 550 °C in UHV, with only small patches of (1 × 1) termination remaining at step edges. Note that panels (e), (f) and (h) show larger areas in order to display the termination change at steps. Sample biases and tunnelling currents are: (a) $U = +3$ V, $I = 0.3$ nA; (b) $U = +3$ V, $I = 0.1$ nA; (c) $U = +2$ V, $I = 0.2$ nA; (d) $U = +2.5$ V, $I = 0.1$ nA; (e, f) $U = +2.5$ V, $I = 0.25$ nA; (g) $U = -2.8$ V, $I = 0.1$ nA; (h) $U = +2.8$ V, $I = 0.1$ nA; (i) $U = -2.8$ V, $I = 0.1$ nA.

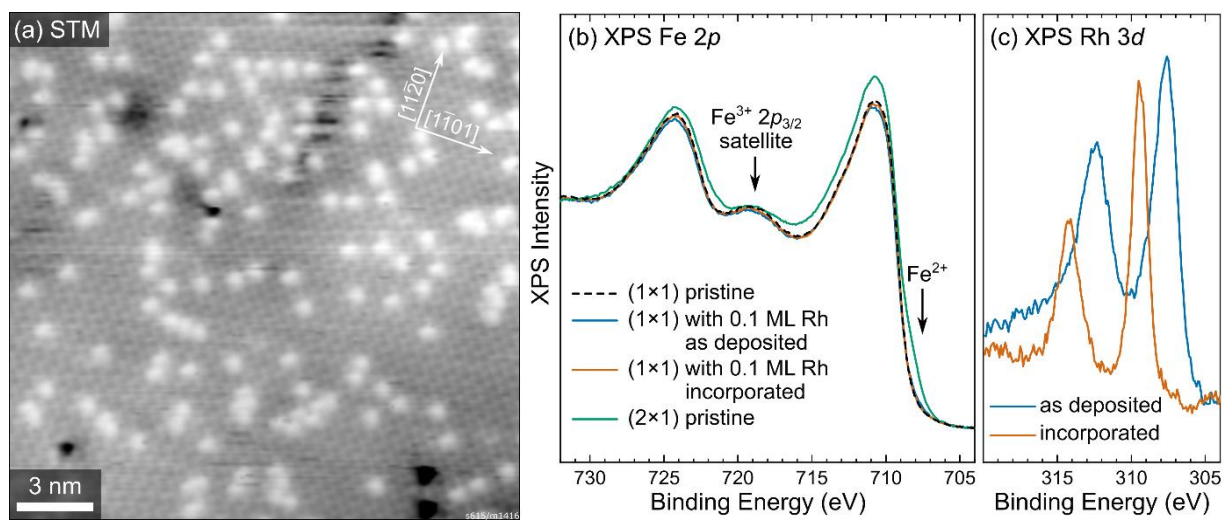

**Figure S2:** 0.1 ML Rh incorporated in α-Fe$_2$O$_3$(1$\bar{1}$02)-(1 × 1). (a) 20 × 20 nm$^2$ STM image ($U_{sample}$ = −2.8 V, $I_{tunnel}$ = 0.1 nA) of the α-Fe$_2$O$_3$(1$\bar{1}$02)-(1 × 1) surface after depositing 0.1 ML Rh, then annealing the sample at 520 °C in 2 × 10$^{-6}$ mbar O$_2$ for 30 m. In contrast to Figure 4 of the main text, no UHV annealing (leading to large, metallic clusters) has been applied. (b,c) The Fe 2$p$ and Rh 3$d$ regions in XPS (Al Kα, 70° grazing emission) for the pristine (1 × 1) surface before Rh deposition (black, dashed), after deposition of 0.1 ML Rh (blue), and corresponding to the STM image in panel (a) (orange). The Fe 2$p$ peak of the as-prepared (2 × 1)-terminated surface is shown for comparison (green). On the (2 × 1) surface, the shoulder at ≈708 eV and the less pronounced Fe 2$p_{3/2}$ satellite at 719 eV indicate the presence of Fe$^{2+}$.[1, 2] These changes are not observed on the (1 × 1) surface even when 0.1 ML Rh are incorporated in the presence of oxygen, suggesting that all iron remains as Fe$^{3+}$.

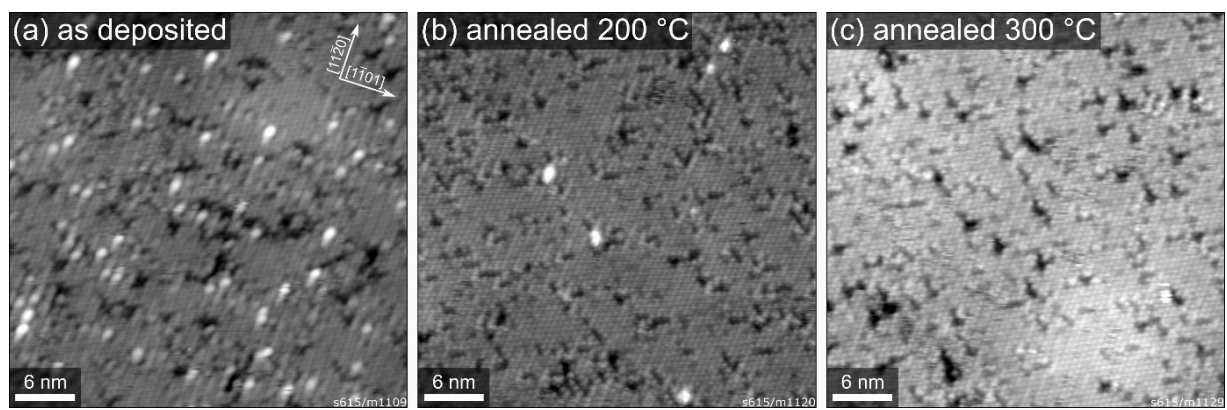

**Figure S3**: STM images corresponding to the XPS data in Figure 2 (b) of 0.025 ML Rh on α-Fe$_2$O$_3$(1$\bar{1}$02)-(2 × 1). Panels (a) and (c) show the same STM images as Figure 1 (c) and (d) at lower magnification. Sample biases and tunnelling currents are: (a) $U = -3$ V, $I = 0.1$ nA; (b) $U = -2$ V, $I = 0.1$ nA; (c) $U = -2.8$ V, $I = 0.1$ nA.

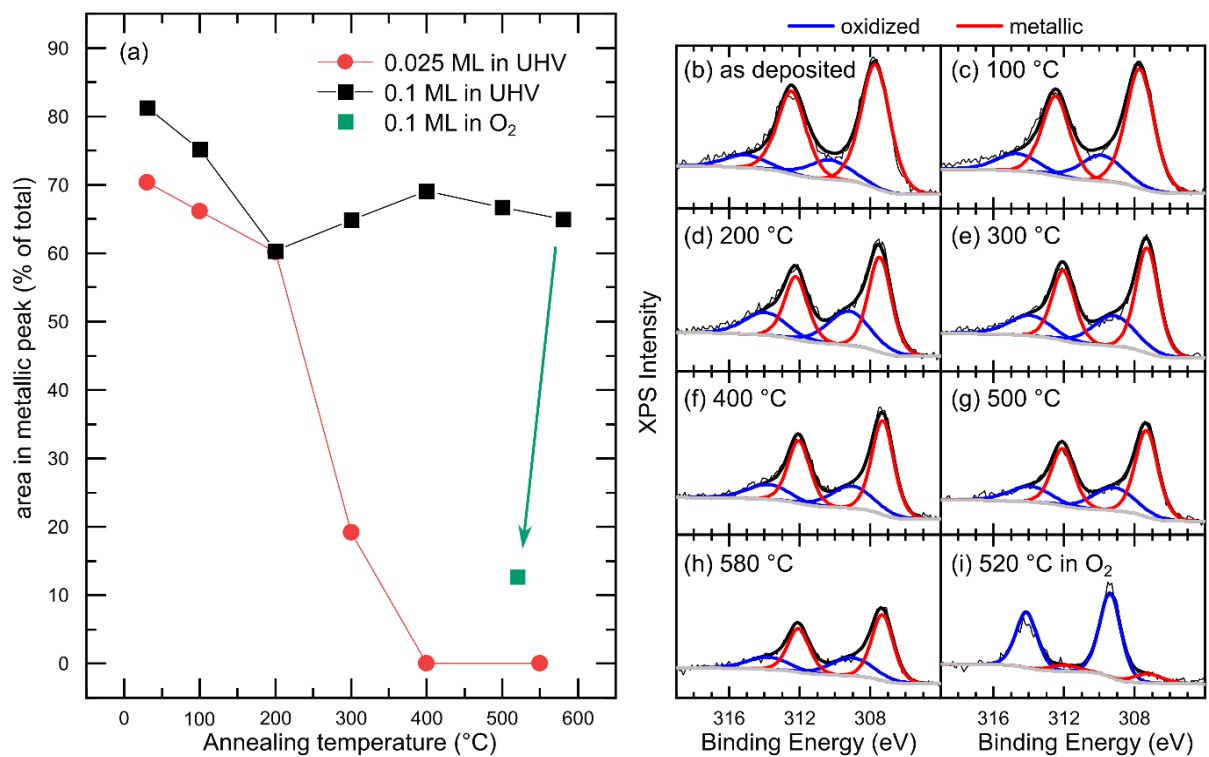

**Figure S4:** Fits of XPS results (Al Kα, 70° grazing emission, pass energy 16 eV) for 0.1 ML Rh on α-Fe$_2$O$_3$(1$\bar{1}$02)-(2 × 1), using the data shown in Figure 4. (a) Area percentages for peak fits to the spectra in Figure 4 (a). Black data points correspond to successive heating steps in UHV, while the green data point corresponds to the final annealing step in oxygen, yielding the STM image in Figure 4 (c). For reference, the data for 0.025 ML Rh from Figure 2 (c) are shown again here (red). Note that in the 0.1 ML case, the peak areas are not good descriptors of the actual ratios between metallic and oxidic Rh because for large clusters, the buried atoms contribute much less signal to XPS. (b-i) Peak fits to the data shown in Figure 4 (a), using two components for Rh 3$d_{5/2}$ as described for Figure 2 in the main text. For the final spectrum in panel (i), the sample was annealed for 1 h at 520 °C in a background of 2 × 10$^{-6}$ mbar O$_2$.

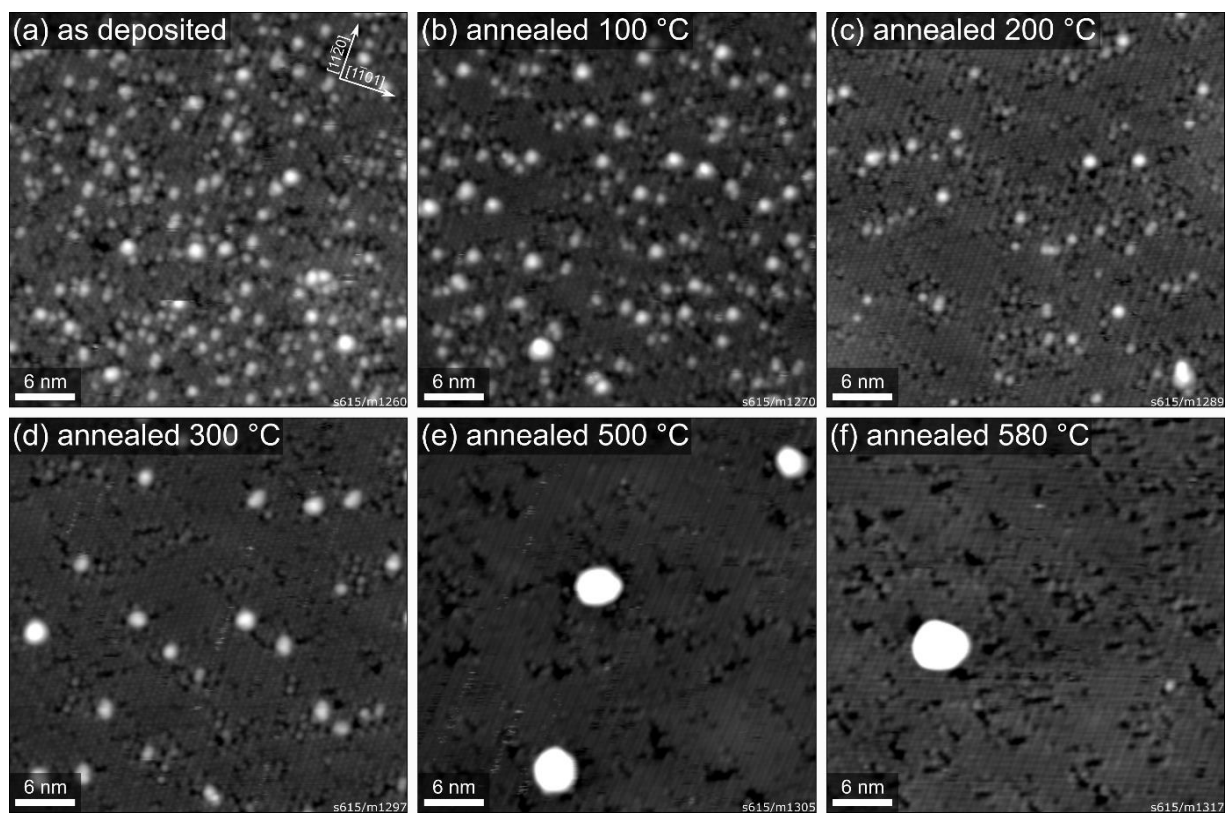

**Figure S5:** STM images corresponding to the XPS data in Figure 4 (a) of 0.1 ML Rh on α-Fe$_2$O$_3$(1$\bar{1}$02)-(2 × 1), with annealing steps performed in UHV. Panel (f) is the same image as shown in Figure 4 (b). Sample biases and tunnelling currents are: (a) $U = -3$ V, $I = 0.1$ nA; (b) $U = -3$ V, $I = 0.1$ nA; (c) $U = -2.5$ V, $I = 0.1$ nA; (d) $U = -3$ V, $I = 0.1$ nA; (e) $U = -3$ V, $I = 0.1$ nA; (f) $U = -2.5$ V, $I = 0.1$ nA.

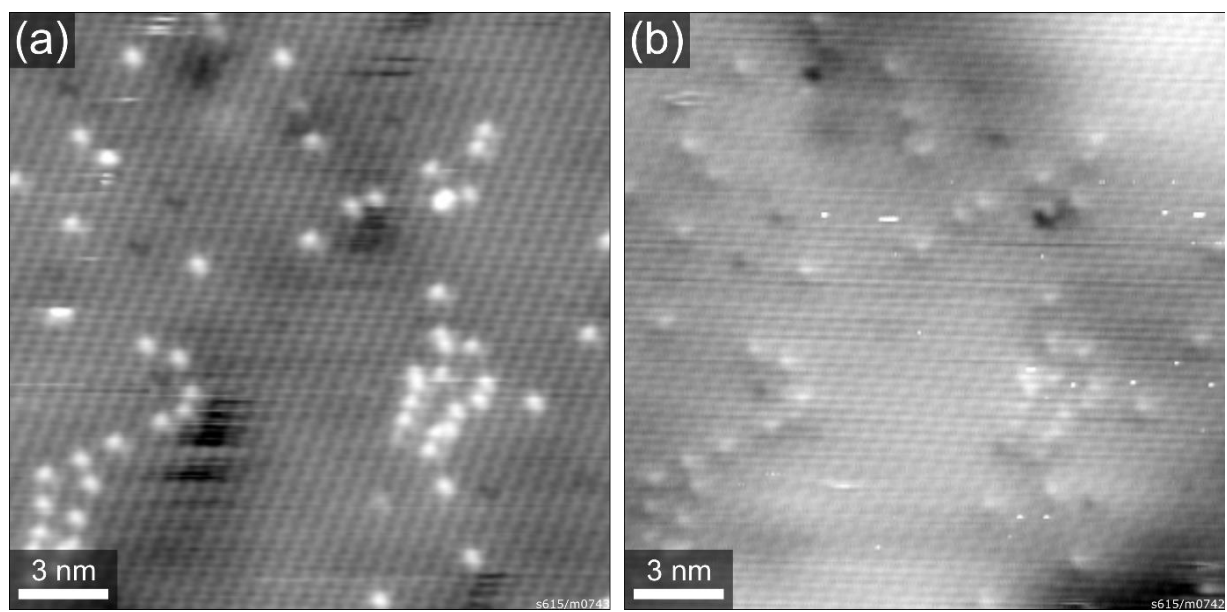

**Figure S6:** STM images of 0.025 ML Rh on α-Fe$_2$O$_3$(1$\bar{1}$02)-(1 × 1). Both images were taken after depositing 0.025 ML Rh on the pristine α-Fe$_2$O$_3$(1$\bar{1}$02)-(1 × 1) surface, followed by annealing at 500 °C for 15 min. The image in panel (a) is the same as shown in the main manuscript in Figure 1(b) ($U_{sample}$ = −2.8 V, $I_{tunnel}$ = 0.1 nA). (b) shows the same area with positive sample bias ($U_{sample}$ = +2.8 V, $I_{tunnel}$ = 0.1 nA).